\documentclass[letterpaper,USenglish,authorcolumns,numberwithinsect]{lipics-v2019}

\newcommand{\hide}[1]{}
\hideLIPIcs
\nolinenumbers

    \newdimen\origiwspc
    \newdimen\origiwstr

\usepackage{mathtools} 
\usepackage[export]{adjustbox} 
\usepackage [lined,boxed,ruled]{algorithm2e} 
\usepackage{stmaryrd} 

\newcommand\vartextvisiblespace[1][.5em]{%
  \makebox[#1]{%
    \kern.07em
    \vrule height.3ex
    \hrulefill
    \vrule height.3ex
    \kern.07em
  }
}
\newcommand{\smallunderscore}{\texttt{\vartextvisiblespace[.7em]}}
\newcommand{\gbar}{\: | \:}
\newcommand{\gvee}{\: \vee \:}
\newcommand{\gdot}{\: \bullet \:}

\DeclareRobustCommand{\DefVar}[1]{\emph{#1}}
\hide{
        \newtheorem{proposition}{Proposition}
        \newtheorem{theorem}{Theorem}
        
        \newtheorem{lemma}{Lemma}
        \newtheorem{definition}{Definition}
        \newtheorem{corollary}{Corollary}
}

\newcommand{\overbar}[1]{\mkern 4.5mu\overline{\mkern-4.5mu#1\mkern-1.5mu}\mkern 1.5mu}

\title{Detecting Opportunities for Differential Maintenance of  Extracted Views}

\author{Besat Kassaie}
{David R. Cheriton School of Computer Science, University of Waterloo, Waterloo, Ontario, Canada, N2L 3G1}{bkassaie@uwaterloo.ca}{}{}

\author{Frank Wm. Tompa}
{David R. Cheriton School of Computer Science, University of Waterloo, Waterloo, Ontario, Canada, N2L 3G1}{fwtompa@uwaterloo.ca}{}{}

\authorrunning{B. Kassaie and F. W. Tompa}
\Copyright{Besat Kassaie and Frank Wm. Tompa}



\begin{document}

\maketitle

\begin{abstract}
 Semi-structured and unstructured data management is challenging, but many of the problems encountered are analogous to problems already addressed in the relational context. In the area of information extraction, for example,
 the shift from engineering \textit{ad hoc}, application-specific extraction rules towards using expressive languages such as \emph{CPSL} and \emph{AQL} creates opportunities to propose solutions that can be applied to a wide range of extraction programs. In this work we focus on \emph{extracted view maintenance}, a problem that is well-motivated and thoroughly addressed in the relational setting.
 
 In particular, we formalize and address the problem of keeping extracted relations consistent with source documents that can be arbitrarily updated. We formally characterize three classes of document updates, namely those that are \textit{irrelevant}, \textit{autonomously computable}, and \textit{pseudo-irrelevant} with respect to a given extractor. Finally, we propose algorithms to detect pseudo-irrelevant document updates with respect to extractors that are expressed as \emph{document spanners}, a model of information extraction inspired by \emph{SystemT}.

\end{abstract}

\section{Introduction}

Designing new languages and extraction platforms~\cite{DBLP:conf/tipster/AppeltO98,cunningham1999jape,DBLP:conf/icde/ReissRKZV08, DBLP:conf/vldb/ShenDNR07}, choosing an appropriate algorithmic approach respecting the domain and the syntactic and semantic properties of anticipated data sources and outputs~\cite{DBLP:conf/emnlp/RitterCME11}, facilitating the incorporation of human knowledge in algorithm design~\cite{DBLP:conf/sigmod/ChaiVDN09}, and  adapting existing extractors to deal with new documents  added to the system~\cite{DBLP:conf/icde/ChenDYR08} cover the significant part of recent research that has been done in this area. In all these efforts, the major goal is to cover the myriad ways that a relationship might be expressed in text.

Despite many technical differences, all proposed extraction approaches share a subtle and important assumption, which we call ``fading attachment.'' The flow of information  between the three main components of information extraction---source documents, the extraction program, and the extracted relations---is maintained during the development period but evaporates once the extraction program reaches a satisfactory level of accuracy and robustness. Once deployed, the information extraction process ignores the relationship between the contents of the source documents and the extracted relations.
 
We observe that the fading attachment assumption is inappropriate in many applications.  Extracted relations might be modified due to privacy concerns~\cite{DBLP:conf/doceng/KassaieT19} or for data cleaning purposes~\cite{IlyasChu2019}, but thereafter they are inconsistent with the contents of the source document. On the other hand, source documents might also be modified, perhaps for versioning purposes or to accommodate updates that reflect the most recent data; but again the extracted relations become inconsistent with the content in the source documents.
 
Instead we consider an extracted relation to be a materialized view of the document corpus. From this perspective, updating extracted relations resembles the classical view update problem for relational databases~\cite{DBLP:conf/vldb/DayalB78}, and keeping extracted relations in sync with the document corpus resembles the problem of maintaining materialized views~\cite{Gupta:1999:MVT:310709}.  The extracted view update problem has been introduced and formalized elsewhere~\cite{DBLP:conf/doceng/KassaieT19}, and in this paper we are interested in the latter problem, i.e., extracted view maintenance. 

The natural way to reflect changes in source documents is to wipe out any extracted relations and repeat the extraction process.  Although this approach guarantees the preservation of consistency between the source text and the extracted relations,
as in the relational database context, extracting relations from scratch can be costly. For instance, in some applications where updates to source documents occur frequently, extraction time might be a bottleneck or, in a distributed setting in which extracted relations and source documents reside in different physical sites, the communication cost for repeatedly transferring  newly extracted relations might be significant. Thus, avoiding re-extraction is sometimes highly desirable.

The problem has been studied extensively in the relational database setting. Based on the requirements of target applications and the nature of view updates, proposed solutions range from recomputing views from scratch to detecting irrelevant and autonomously computable updates~\cite{DBLP:journals/tods/BlakeleyCL89} and to updating views differentially~\cite{DBLP:conf/sigmod/GuptaMS93,DBLP:conf/dbpl/KawaguchiLMR97} or only as needed~\cite{DBLP:conf/sigmod/ColbyGLMT96,DBLP:conf/icde/ZhouLGD07}. Other optimization techniques can also be adopted from relational databases~\cite{DBLP:conf/icde/ReissRKZV08}, including the materialization of partially extracted views (which would also need to be maintained, of course). In fact, we hypothesize that any of the proposed solutions in the relational setting can be adapted to the extracted view maintenance problem. 
 
However, due to the diverse range of extraction techniques and \textit{ad hoc} document updates, tackling the problem of extracted view maintenance introduces new challenges. Given a collection of documents $\mathbb{D}$, a set of extraction programs $\mathbb{E}$, a corresponding set of extracted relations $\mathbb{R}$, and an instance of a document update specification $U$, we study conditions under which we can apply $U$ to members of $\mathbb{D}$ and apply corresponding updates to members of $\mathbb{R}$ without recomputing the revised extracted relations from scratch (Figure~\ref{fig:docDB}).
That is, we wish to translate updates over documents into differential updates over extracted relations. Thus, in this paper:

\begin{figure}[t]
 \centering
\includegraphics[width=0.5\columnwidth]{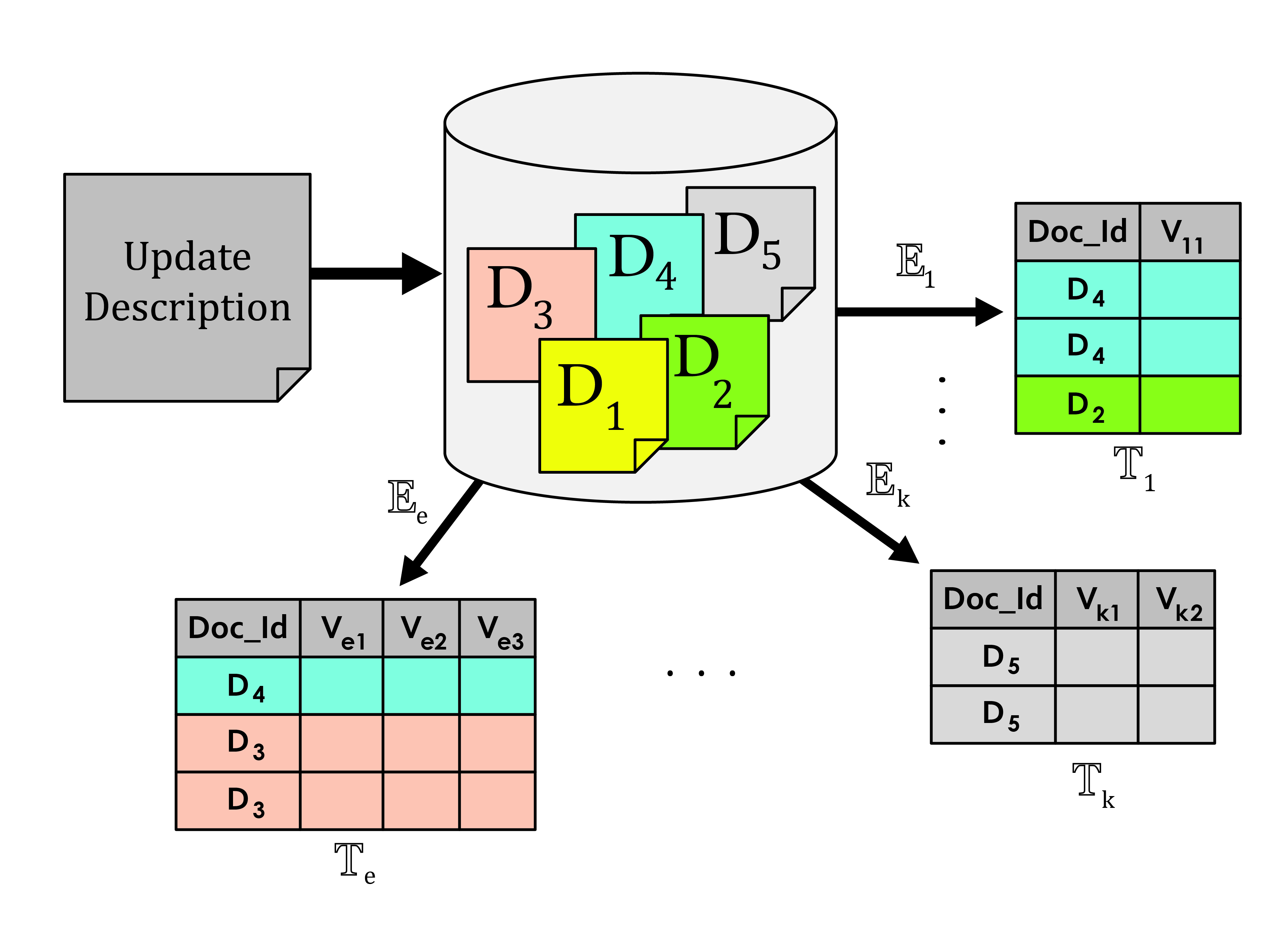}
\caption{Extraction system that supports updates to all members of a document database.} 
\label{fig:docDB}
\end{figure}

\begin{enumerate}[(a)]
\item We introduce the extracted view maintenance problem.
\item We propose a match-and-replace document update model.
\item We formalize three categories of document updates for which we can preserve consistency without repeating the extraction process: \textit{irrelevant}, \textit{autonomously computable}, and \textit{pseudo-irrelevant updates}. 
\item We propose algorithms to determine whether an update is pseudo-irrelevant with respect to extractors expressed as document spanners, a formalism that models the basis of the SystemT extraction system~\cite{DBLP:conf/icde/ReissRKZV08}.
 \end{enumerate}
 
 \section{Preliminaries}
In order to develop specific algorithms, we assume that extracted views are defined using SystemT, an information extraction platform that benefits from relational database concepts to deal with text data sources~\cite{DBLP:conf/icde/ReissRKZV08}\footnote{How to maintain extracted views efficiently should also be investigated using other extraction languages, such as JAPE~\cite{cunningham1999jape}.}. SystemT models each document as a single string and populates relational tables with \textit{spans}, directly extracted from the input document. With SystemT users encode extractors with a SQL-like language, i.e., AQL,  to manipulate tables. AQL offers operators to work directly on text or on the extracted tables (standard relational operators  that accept span predicates).

The underlying principles adopted by SystemT have been formalized as \textit{document spanners} by Fagin, et al.~\cite{DBLP:journals/jacm/FaginKRV15}. Most of the material in this section has been introduced in that work, which contains additional details.

Let $\Sigma$ be a finite alphabet and $D$ be a (finite) document over $\Sigma$, i.e., $D \in \Sigma^{*}$. A \DefVar{span} of $D$, denoted $[i, j \rangle$ ($1 \leq i \leq j \leq |D|+1$), specifies the start and end offsets of a substring in $D$, which is in turn denoted $D_{[i, j \rangle}$, and extends from offset $i$ through offset $j-1$. If $i=j$, this denotes an empty span at offset $i$. Spans $s_1 = [i_1, j_1 \rangle$  and $s_2 = [i_2, j_2 \rangle$ are identical if and only if $i_1=i_2$ and $j_1=j_2$; they overlap if $i_1 \leq i_2 < j_1$ or $i_2 \leq i_1 < j_2$. 
Regular expressions extended using variables chosen from a set $V$ are called \DefVar{regular expressions with capture variables}, defined by $\gamma$ in the grammar $G_S(\Sigma,V)$ as follows:
\[\gamma\: :=\: \emptyset \gbar\epsilon \gbar \sigma \gbar (\gamma \gvee \gamma) \gbar (\gamma \gdot \gamma) \gbar (\gamma)^{*} \gbar x\{\gamma\}\]
\noindent where $\sigma \in \Sigma$ and $x \in V$.
The use of a sub-expression of the form $x\{g\}$ is to denote that whenever the regular expression matches a string, sub-strings matched by $g$ are to be \emph{marked by the capture variable $x$}. If $E$ is a regular expression with capture variables, then we denote the set of capture variables in $E$ as $\mathit{SVars}(E)$. We use $G_S$ in place of $G_S(\Sigma,V)$ whenever $\Sigma$ and $V$ are immaterial or understood from the context. We also allow regular expressions with capture variables to be written without parentheses that can be inferred based on priority of operations~\cite{DBLP:books/daglib/0016921}.

Applying an information extractor to a document $D$ produces a \DefVar{span relation}, i.e., a relation that contains spans of $D$.
To this end, if $E$ is a regular expression with capture variables, it specifies a \DefVar{document spanner}, denoted $\llbracket E \rrbracket$, which is a function mapping strings over $\Sigma^{*}$ to span relations. In particular, for a given document $D$, the spanner specified by $E$ produces a span relation $\llbracket E \rrbracket (D)$ in which there is one column for each variable from $V$ appearing in $E$, each row corresponds to a matching of $E$ against $D$ when the variables are ignored, and the value in a row for the column corresponding to $x \in V$ is the span marked by $x$. To ensure that the extracted relation is in first-normal form with no null values, we restrict our attention to a specific class of document spanners, namely \DefVar{functional document spanners}, that assign exactly one span to each variable for all produced rows, regardless of the input document $D$.

Let $\Sigma$ be the set of Latin alphanumeric, punctuation and the space characters (the last represented by {\smallunderscore}), and let $d$ denote a digit. 
Applying \[{\gamma}_{phone}={\Sigma}^{*} \bullet tn\{(0\bullet 1\vee \texttt{1}\vee +\bullet 1)\bullet\texttt{-}\bullet ac\{d\bullet d\bullet d\}\bullet\texttt{-}\bullet d\bullet d\bullet d\bullet\texttt{-}\bullet sc\{d\bullet d\bullet d\bullet d\}\}\bullet {\Sigma}^{*}\]
to the document in Figure~\ref{fig:inputtext}  results in the span relation in Figure~\ref{fig:spannerPhonenumbers}. 
\begin{figure}
 \centering
 \renewcommand{\arraystretch}{1} 
 \setlength\tabcolsep{0.5pt} 
\begin{tabular}{ccccccccccccccccccccccccccccccccccccccccc}
F&o&r&{\smallunderscore}&i&n&f&o&r&m&a&t&i&o&n&{\smallunderscore}&o&n&{\smallunderscore}& C&O&V&I&D&-&1&9&{\smallunderscore}&,&{\smallunderscore}&c&a&l&l&{\smallunderscore}&u&s&{\smallunderscore}&a&t&{\smallunderscore}\\
\hline
\tiny1&\tiny2&\tiny3&\tiny4&\tiny5&\tiny6&\tiny7&\tiny8&\tiny9&\tiny10&\tiny11&\tiny12&\tiny13&\tiny14&\tiny15&\tiny16&\tiny17&\tiny18&\tiny19&\tiny20&\tiny21&\tiny22&\tiny23&\tiny24&\tiny25&\tiny26&\tiny27&\tiny28&\tiny29&\tiny30&\tiny31&\tiny32&\tiny33&\tiny34&\tiny35&\tiny 36&\tiny37&\tiny38&\tiny39&\tiny40&\tiny41\\
\hline
1&-&8&3&3&-&7&8&4&-&4&3&9&7&{\smallunderscore}&,&{\smallunderscore}& s&p&e&c&i&f&i&c&{\smallunderscore}&t&o&{\smallunderscore}& y&o&u&r&{\smallunderscore}&p&r&o&v&i&n&c\\
\hline
\tiny 42&\tiny43&\tiny44&\tiny45&\tiny46&\tiny47&\tiny48&\tiny49&\tiny50&\tiny51&\tiny52&\tiny53&\tiny54&\tiny55&\tiny56&\tiny57&\tiny58&\tiny59&\tiny60&\tiny61&\tiny62&\tiny63&\tiny64&\tiny65&\tiny66&\tiny67&\tiny68&\tiny69&\tiny70&\tiny71&\tiny72&\tiny73&\tiny74&\tiny75&\tiny76&\tiny77&\tiny78&\tiny79&\tiny80&\tiny81&\tiny82\\
\hline
e&{\smallunderscore}&a&t&{\smallunderscore}&+&1&-&8&6&7&-&9&7&5&-&5&7&7&2&{\smallunderscore}&o&r&{\smallunderscore}&4&0&3&-&6&4&4&-&4&5&4&5&{\smallunderscore}&.\\
\hline
\tiny83&\tiny84&\tiny85&\tiny86&\tiny87&\tiny88&\tiny89&\tiny90&\tiny91&\tiny92&\tiny93&\tiny94&\tiny95&\tiny96&\tiny97&\tiny98&\tiny99&\tiny100&\tiny101&\tiny 102&\tiny103&\tiny104&\tiny105&\tiny106&\tiny107&\tiny108&\tiny109&\tiny110&\tiny111&\tiny112&\tiny113&\tiny114&\tiny115&\tiny116&\tiny117&\tiny118&\tiny119&\tiny120
 \end{tabular}
 \caption{A sample input document $D$ for our running example.}
 \label{fig:inputtext}
\end{figure}

\begin{figure}[ht]
 \centering
 \renewcommand{\arraystretch}{1.1} 
 \setlength\tabcolsep{1pt} 
 \begin{tabular}{ccc}
 tn&ac&sc\\
 \hline
 \hline
 $[42,56\rangle$&$[44,47\rangle$&$[52,56\rangle$\\
 \hline
  $[88,103\rangle$&$[91,94\rangle$&$[99,103\rangle$
\end{tabular}
 \caption{The extracted relation $\llbracket{\gamma}_{phone}\rrbracket(D)$, where $D$ is depicted in Figure~\ref{fig:inputtext}.}
 \label{fig:spannerPhonenumbers}
\end{figure}

\begin{definition}Throughout this paper, a functional document spanner used for the purpose of information extraction is called an \DefVar{extraction spanner} or simply an \DefVar{extractor}, the regular expression with capture variables defining it is called an \DefVar{extraction formula}, and the span relation produced for a document is called an \DefVar{extracted relation}.
\end{definition}

\begin{definition}
  A regular expression created by eliminating all the capture variables from an extraction formula $E$ is called the \emph{corresponding Boolean spanner} and is denoted by $B(E)$.  
\end{definition}

In this paper we hypothesize systems that include a document database $\mathbb{D}$ and a set of extractors $\{\mathbb{E}_1, \cdots, \mathbb{E}_e\}$ that run independently over $\mathbb{D}$. The union of span relations produced by an extractor $\mathbb{E}_k$ against the document database is stored in a relation $\mathbb{T}_k$ that includes an additional column to associate each document identifier with the spans for the corresponding span relation.
 These tables serve as materialized views of the document database.

In numerous proofs in this paper, we rely on finding ``witness'' documents that exhibit certain properties:

\begin{definition}
  Given a document $D$ and a property $P$, if $D$ exhibits property $P$ (i.e., we can assert $P(D)$), $D$ is called a \emph{witness for $P$}.
\end{definition}

\noindent Furthermore, we use specially-constructed finite automata to test properties of given spanners. For each automaton, $Q$ represents a finite set of states, $\Sigma$ is the input alphabet, $\delta$ stands for the transition function, $Q_0$ is a set of initial states, and $F$ represents a set of final states.

\section{Document Update Model}
Substring \textit{replacement}, \textit{deletion}, and \textit{insertion} are basic update operations over documents.  A change to the text is typically preceded by some browsing activities or search operations to locate update positions in a target document. In this section we describe the proposed formal model for document update. 

Target points of change in a document are specified using patterns over the input string, expressed as a functional document spanner with precisely one variable. Specifically, an \DefVar{update formula} is an extraction formula for specifying an update, defined by $\gamma$ in the following grammar $G_U(\Sigma,x)$:
\begin{align}
&\gamma\: :=\: (\gamma \gvee \gamma) \gbar (\gamma^{\prime} \gdot \gamma) \gbar (\gamma \gdot \gamma^{\prime}) \gbar x\{\gamma^{\prime}\}\\
&\gamma^{\prime}:=\: \emptyset \gbar \epsilon \gbar \sigma \gbar (\gamma^{\prime} \gvee \gamma^{\prime}) \gbar (\gamma^{\prime} \gdot \gamma^{\prime}) \gbar (\gamma^{\prime})^{*}\label{regex:gamma'}
\end{align} 
\noindent (i.e., where $\gamma^{\prime}$ is a standard, variable-free regular expression).

The functional document spanner that is represented by an update formula $g$ maps every document $D$ to a unary span relation, which we call the \textit{update  relation} and denote as $\llbracket g \rrbracket (D)$. 
When the spanner is used for updating a document $D$, we require that all spans in $\llbracket g \rrbracket (D)$ be mutually disjoint. In this case, sub-strings of $D$ associated with the spans in the update relation are simultaneously replaced by a new value denoted by a constant $A$. 
Because the update relation contains non-overlapping spans, such replacements will be mutually non-interfering.

\begin{definition}
  An instance of an update specification with given update formula $g$ and $A \in \Sigma^{*}$ is called an \emph{update expression} and represented by $\mathit{Repl}(g,A)$. Given a document $D$, if $\llbracket g \rrbracket (D)$ contains no overlapping spans, then applying $\mathit{Repl}(g,A)$ to $D$ produces a new document $\mathit{Repl}(g,A)(D)$ that is identical to $D$ but with every substring in $D$ marked by $x$ in $\llbracket g \rrbracket$ replaced by the string $A$.
\end{definition}

Note that if $A$ is the empty string, then the update results in the deletion of the substrings identified by the spanner; otherwise, wherever the spanner produces an empty span, the replacement, in effect, inserts the string $A$. 
For example, given $D$ as in Figure~\ref{fig:inputtext}, applying  $\mathit{Repl}({\Sigma}^{*}\bullet \texttt{u}\bullet \texttt{s}\bullet${\smallunderscore}$\bullet x\{\epsilon\}\bullet \texttt{a}\bullet \texttt{t}\bullet {\Sigma}^{*}$,\texttt{free}{\smallunderscore}) to $D$ 
inserts  `\texttt{free}{\smallunderscore}', at $[39,39\rangle$.

\subsection{Properties of all update expressions}

\begin{proposition}
  If $g$ is an update formula, then $\llbracket g\rrbracket$ is functional.
\end{proposition}
\begin{proof}
By induction on the height of the \emph{parse} tree for $g$ derived from the root symbol $\gamma$.
\end{proof}

\begin{lemma} \label{lemma:disjFrom}
  Given $\Sigma$ and $V$, let $\bar{\gamma}$ define a restricted form for extraction formulas as follows:
\[
\bar{\gamma} := \gamma^\prime \gbar (\:(\bar{\gamma} \gdot)? \:\: x\{\bar{\gamma}\} \:\: (\gdot \bar{\gamma})?\:)\\
\] 
\noindent where `?' denotes optional and $\gamma^\prime$ is defined in production~(\ref{regex:gamma'}) above.
Every functional extraction formula $E$ based on $G_{S}(\Sigma, V)$ can be rewritten in its \DefVar{normalized form}
$\Delta(E)=\bigvee_{i=1}^{k} E_i$
\noindent where $E_i$ is a formula defined by $\bar{\gamma}$ for $i \in 1..k$. (Note that within each $E_i$, all operands for disjunction and Kleene closure are standard, variable-free regular expressions.)
\end{lemma}

\begin{proof}
By induction on the height of the \emph{expression} tree for $\gamma$.\footnote{An alternative proof can be derived by noting that every extraction formula can be represented by a ``vstk-path union''~\cite{DBLP:journals/jacm/FaginKRV15}.}
\end{proof}
For example, consider the extraction formula
\[E = (a \gvee b)^* \gdot X\{(Y\{a\} \gvee Y\{a \gdot b\}) \gdot a\} \gdot Z\{b \gvee (b \gdot a)\}\]
\noindent The normalized form for $E$ is
\begin{align*}
    \Delta(E) = &(a \gvee b)^* \gdot X\{Y\{a\} \gdot a\} \gdot Z\{b \gvee (b \gdot a)\} \gvee \\
    &(a \gvee b)^* \gdot X\{Y\{a \gdot b\} \gdot a\} \gdot Z\{b \gvee (b \gdot a)\}
\end{align*}

In short, to normalize a formula, all disjunctions that have capture variables in their disjuncts\footnote{Because the formulas are functional, if a capture variable appears in one disjunct, it must appear in all disjuncts.} can be ``pulled up'' over concatenations and other capture variables in the expression tree to create separate disjuncts at the outermost level of the formula. 
\begin{lemma}
  Given an extraction formula $E$ with $v=|V|$ capture variables and at most $d$  disjuncts per capture variable, $k \leq d^v$ in the normalized form $\Delta(E)$.\footnote{For all practical purposes, this is a polynomial blowup in expression size.}
\end{lemma} 
\begin{proof}
By induction on $v$.
\end{proof}

\begin{corollary} \label{Cor:disjunctForm}
  Every  update formula $g$ can be rewritten as $\Delta(g)$, a disjunction of the form
$\bigvee_{i=1}^{k} U_i$
\noindent where $U_i$ is a formula defined by $(\gamma^{\prime}  \gdot)?\:\:  x\{\gamma^{\prime}\} \:\: (\gdot  \gamma^{\prime})?$ for $i \in 1..k$, $\gamma^{\prime}$ is defined by production~(\ref{regex:gamma'}), $k \leq d$, and $d$ is the number of disjuncts including the capture variable in $g$.
\end{corollary}

\subsection{Unrestricted Update Spanners}\label{DisjointUpdateSpanner}

As noted earlier, we require that an update spanner produces no overlapping spans. 
\begin{definition}
  An update spanner is \emph{unrestricted} if, for every input document, the spans marked by the capture variable $x$ are pairwise identical\footnote{Note that even though a span relation is a set, not a bag, the same span might be marked through more than one match to the update formula.} or non-overlapping, i.e., there does not exist a witness for overlapping spans. 
\end{definition}

\begin{figure}[bt]
\centering
\includegraphics[width=0.7\columnwidth]{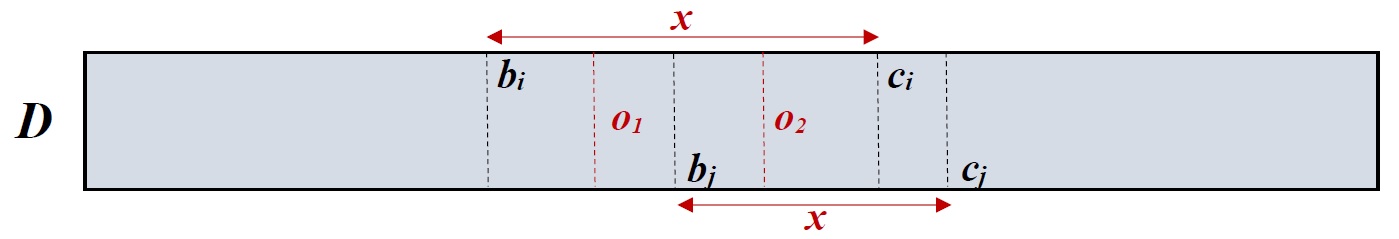}
\caption{$D$ is a witness for overlapping spans for $\llbracket g \rrbracket$, where $s_i=[b_i,c_i \rangle $ and $s_j=[b_j,c_j \rangle $ are spans marked by \textit{x} when matched by the $i^{\tiny th}$ and $j^{\tiny th}$ disjuncts (not necessarily distinct) of $\Delta(g)$. Offset $o_1$ falls within $s_i$ but not $s_j$, and $o_2$ falls within both $s_i$ and $s_j$.} 
\label{fig:updates}
\end{figure}

To determine whether the set of witnesses for overlapping spans is provably empty (Figure~\ref{fig:updates}),

we first normalize $g$ and then construct the automaton $\mathcal{M}_{g}$ that matches $\Delta(g)$ using standard techniques~\cite{DBLP:books/daglib/0016921}.
Let   
$U_i=\gamma^{\prime}_{L_i} \gdot x\{\gamma^{\prime}_{C_i}\} \gdot \gamma^{\prime}_{R_i}$
represent the $i^{\tiny th}$ disjunct of $\Delta(g)$.
Then, for each disjunct $U_i$, let the finite automaton for every (variable-free) sub-expression $\gamma^{\prime}_{L_i}$, $\gamma^{\prime}_{C_i}$, and $\gamma^{\prime}_{R_i}$ be represented by $\mathcal{M}_{L_i}, \mathcal{M}_{C_i}$, and $\mathcal{M}_{R_i}$ respectively. Formally:
$\mathcal{M}_{L_i}=<Q_{L_i},\Sigma_{L_i}, \delta _{L_i}, Q_{0_{L_i}}, F_{L_i}>$,
$\mathcal{M}_{C_i}=<Q_{C_i},\Sigma _{C_i}, \delta _{C_i}, Q_{0_{C_i}}, F_{C_i}>$, and
$\mathcal{M}_{R_i}=<Q_{R_i},\Sigma_{R_i}, \delta _{R_i}, Q_{0_{R_i}}, F_{R_i}>.$

$\mathcal{M}_{U_i}$ is constructed by applying the standard concatenation operator to $\mathcal{M}_{L_i}$, $\mathcal{M}_{C_i}$, and $\mathcal{M}_{R_i}$; that is, 
$
 \mathcal{M}_{U_i}=\mathcal{M}_{L_i} \gdot \mathcal{M}_{C_i}  \gdot \mathcal{M}_{R_i} 
$.

Finally, we construct $\mathcal{M}_{g}=<Q_{g},\Sigma_{g}, \delta _{g}, Q_{0_{g}}, F_{g}>$ by applying the standard union operator to the $\mathcal{M}_{U_{i}}$ machines. Then $Q_g=  Q_L \cup Q_C \cup Q_R$ in which $Q_L$ is the union of states in $Q_{L_{i}}$, $Q_C$ is the union of states in $Q_{C_{i}}$, and $Q_R$ is the union of states in $Q_{R_{i}}$. 
Given $a \in \Sigma_{g}$, $Q \subseteq Q_{g}$, and $q \in Q_{g}$, let $\mathcal{C}(Q,q,a)$ denote the predicate ``$q \in Q \wedge \delta_{g}(q,a) \in Q$'' and $\mathcal{C}(Q_1,q,a,Q_2)$ denote the predicate ``$q \in Q_1 \wedge \delta_{g}(q,a) \in Q_2$.''

We build the following automaton to identify the set of witnesses for overlapping spans for a given $\Delta(g)$ represented by $\mathcal{M}_g$. Each state encodes four properties: the state of matching for each of two (not necessarily distinct) disjuncts in $\Delta(g)$, whether or not the matched spans are different, and whether or not the spans overlap.\footnote{For simplicity, we refer to the last two dimensions of each state as if they were variables named $v$ and $w$, respectively.}
\begin{description}
\item [$\mathcal{M}_{\Xi}$] $=<Q_{\Xi},\Sigma_{\Xi}, \delta _{\Xi}, Q_{0_{\Xi}},F_{\Xi}>$ where
\item[ ] $\Sigma_{\Xi}=\Sigma_{\mathcal{M}_{g}},$ 	 
\item[ ] $Q_{\Xi}=Q_{g} \times Q_{g} \times \{T,F\} \times \{T,F\},$ 
\item[ ] $Q_{0_{\Xi}}=\{ (q_{i}, q_{j},F,F) \gbar  q_{i} \in Q_{0_{g}} \wedge  q_{j} \in Q_{0_{g}}\},$
\item[ ] $F_{\Xi}=\{ (q_{i}, q_{j},T,T) \gbar q_{i} \in F_{g} \wedge q_{j} \in F_{g}\}$ 
\item[ ] $\delta_{\Xi}((q_{i} \times q_{j} \times v \times w), a) =  
   \begin{dcases*} 
   (\delta_{g}({q_{i}},a),\delta_{g}({q_{j}},a),T,w)&if {$\mathcal{C}(Q_C,q_i,a) \wedge \mathcal{C}(Q_L,q_j,a)$}\\
   (\delta_{g}({q_{i}},a),\delta_{g}({q_{j}},a),T,w) &if {$\mathcal{C}(Q_C,q_i,a) \wedge \mathcal{C}(Q_R,q_j,a)$}\\
   (\delta_{g}({q_{i}},a),\delta_{g}({q_{j}},a),v,T) &if {$\mathcal{C}(Q_C,q_i,a) \wedge \mathcal{C}(Q_C,q_j,a)$}\\
   (\delta_{g}({q_{i}},a),\delta_{g}({q_{j}},a),T,T) &if {$\mathcal{C}(Q_C,q_i,a) \wedge  \mathcal{C}(Q_L,q_j,a,Q_R)$}\\
   (\delta_{g}({q_{i}},a),\delta_{g}({q_{j}},a),v,w) & otherwise.
   \end{dcases*}  $
\end{description}

\begin{proposition}\label{prop:disjointUpdate}
  $L(\mathcal{M}_{\Xi})= \{D | D $ is witness for overlapping spans for $\llbracket g \rrbracket\}$.
\end{proposition}
\begin{proof}
   We first show that if $D$ is a witness for overlapping spans for $\llbracket g \rrbracket$ then $D \in  L(\mathcal{M}_{\Xi})$.
Being a witness implies that 
$D$ can be matched in at least two different ways: using $\mathcal{M}_i$ and $\mathcal{M}_j$.
Let spans $s_i$ and $s_j$ be marked by $x$ in $\mathcal{M}_i$ and $\mathcal{M}_j$, respectively.
If they  are overlapping, there exist two offsets $o_1$ and $o_2$ (not necessarily distinct) as defined in Figure~\ref{fig:updates}.
$s_i$ and $s_j$ cannot both be empty: if they were, they would be either identical or disjoint by definition.
\begin{enumerate}[(i)]
    \item If one of the spans, say $s_i$, is not empty and the other, say $s_j$, is empty, let $o_2$ be an offset that falls within both spans and let $a$ be the symbol at offset $o_2$ in $D$. Because $o_2$ falls in the span matched by $\mathcal{M}_{C_i}$ for $\mathcal{M}_i$, reading $a$ causes a transition to some (other) state in $M_{C_i}$ in $\mathcal{M}_i$. However, because $s_j$ is empty, there are only epsilon transitions between the initial state(s) and final state(s) of $\mathcal{M}_{C_j}$. Therefore reading $a$ at $o_2$ causes a transition from some state in $\mathcal{M}_{L_j}$ to some state in $M_{R_j}$ for $\mathcal{M}_j$. The fourth alternative in the definition of the transition function in $\mathcal{M}_{\Xi}$ sets $v=w=T$, and further transitions will eventually lead to a final state. 
    \item If $s_i$ and $s_j$ are both non-empty, then reading a symbol at $o_1$ will cause a transition from some state in $\mathcal{M}_{C_i}$ to some (other) state in $\mathcal{M}_{C_i}$ while not making such a transition in $\mathcal{M}_{C_j}$ (i.e., either wholly within $\mathcal{M}_{L_j}$ or $\mathcal{M}_{R_j}$). However, reading the symbol at $o_2$ will cause a transition from some state in $\mathcal{M}_{C_i}$ to some (other) state in $\mathcal{M}_{C_i}$ as well as from some state in $\mathcal{M}_{C_j}$ to some (other) state in $\mathcal{M}_{C_j}$. The first of these sets $v=T$, and the second sets $w=T$. Thus when the input is exhausted, $\mathcal{M}_{\Xi}$ will be in a final state.
\end{enumerate}
Second we show that  $D \in L(\mathcal{M}_{\Xi})$ implies that $D$ is a witness for overlapping spans for $\llbracket g \rrbracket$. By construction, if $\mathcal{M}_{\Xi}$ accepts an input, it corresponds to starting in an initial state and ending in a final state of $\mathcal{M}_g$. Furthermore, marking both $w=T$ and $v=T$ necessitates that the input contains two offsets $o_1$ and $o_2$ (not necessarily distinct) as defined in Figure~\ref{fig:updates}. 
\end{proof}
\begin{corollary}\label{cor:unrestricted}
  Let $g$ be an update formula and construct $\mathcal{M}_{\Xi}$ as above. If  $\mathit{min}(\mathcal{M}_{\Xi}) = \emptyset$\footnote{The \emph{min} function represents standard state minimization.} then $\llbracket g \rrbracket$ is an unrestricted update spanner.
\end{corollary}	
\section{Irrelevant and Autonomously Computable Updates}
As defined above, applying an update expression $\mathit{Repl}(g,A)$ to an input document $D$, where $g$ specifies an unrestricted update spanner, returns a new document $D'$ in which the contents of each span identified by $\llbracket g \rrbracket$ is replaced by the string $A$. Given an update expression and an extraction spanner, we wish to determine, for all potential input documents, whether the extracted materialized view can be kept consistent with the updated source documents without running the extractor after updating the documents in the database. This problem is  similar to filtering out irrelevant updates  or applying updates autonomously to relational materialized views~\cite{Blakeley:1999:EUM:310709.310739}. 

\begin{definition}\label{Def: Irrelevant}
  An update expression $\mathit{Repl}(g,A)$ is \DefVar{irrelevant} with respect to an extractor $\llbracket E \rrbracket$ if for every input document, applying $\llbracket E \rrbracket$ to $\mathit{Repl}(g,A)(D)$ produces a span relation that is identical to applying $\llbracket E \rrbracket $ to $D$. That is, if $D' \! = \! \mathit{Repl}(g,A)(D)$, then $\llbracket E \rrbracket (D') = \llbracket E \rrbracket(D)$.
\end{definition}

If an update expression is relevant with respect to an extractor, it may be that the modification to the extracted relation can be computed without re-running the extractor.

\begin{definition}\label{Def:autonomous}
  An update expression $\mathit{Repl}(g,A)$ is \DefVar{autonomously computable} with respect to an extractor $\llbracket E \rrbracket$ if for every input document, applying $\llbracket E \rrbracket$ to $\mathit{Repl}(g,A)(D)$ can be computed from the update expression, the update relation, the extraction formula that defines the extraction spanner, and the extracted relation.\footnote{Autonomous computability for updates is analogous to determinacy~\cite{DBLP:journals/tods/NashSV10} for queries.} 
\end{definition}

There is an important distinction between the problems of updating traditional relational views and updating materialized extractions. Span relations contain pairs of offsets from input documents, not document content. Thus a span relation might be affected by an update even if the replaced text is not within an extracted span. In particular, replacing a string of one length by a string of another length somewhere in the document will cause a span somewhere else in the document to shift, even if the content of that span is unaffected.

More specifically, given a document $D$ and the corresponding updated document $D'$, if span $S$ in $D$ is disjoint from all spans produced by the unrestricted update spanner $\llbracket g \rrbracket$, let $\mathit{shift}(g,A)(S)$  represent the corresponding span in $D'$, i.e., the new location of the content of $S$ in $D'$. $\mathit{shift}(g,A)(S)$ is shifted from $S$ by an amount that is dependent on the length of $A$ and the lengths of all spans in the update relation that precede $S$ in $D$, 
as captured by Algorithm~\ref{alg-Shift}.
\begin{figure}[ht]
 \centering
 \begin{minipage}{.9\textwidth}
\begin{algorithm}[H]
    \SetKwInput{Pre}{Precondition}
    
	\KwIn{update relation $R_U$, $A$,  span $S= [i,j \rangle$}
	\KwOut{span $S'= [i',j' \rangle=\mathit{shift}(g,A)(S)$}
	\Pre{$R_U$ contains no duplicates and no span that overlaps $S$ or any other span in $R_U$}
$\mathit{shift} \gets 0$\;
 
	\For {tuple  $ [m,n \rangle \in R_U$ }
	{  
		\If{m $<$ i} {
	  $\mathit{shift} \gets \mathit{shift}+(n-m)-\mathit{length(A)}$}
 	}
	\Return $[i-\mathit{shift},j-\mathit{shift} \rangle$
		\caption{Shift Algorithm.}
	\label{alg-Shift}
\end{algorithm} 
 \end{minipage}
\end{figure}

\begin{definition}\label{Def:pseudo-irrelevant}
  Update expression $\mathit{Repl}(g,A)$ is \DefVar{pseudo-irrelevant} with respect to an extraction spanner $\llbracket E \rrbracket$ if for every input document, applying $\llbracket E \rrbracket$ to $\mathit{Repl}(g,A)(D)$ produces a span relation that is identical to applying $\llbracket E \rrbracket$ to $D$ except to replace each span $S$ by $\mathit{shift}(g,A)(S)$. That is, if $D' = \mathit{Repl}(g,A)(D)$, then $\llbracket E \rrbracket (D') = \{S' \;|\; \exists\; S \in \llbracket E \rrbracket(D)$ such that $S'=\mathit{shift}(g,A)(S) \}$.
\end{definition}

Thus, a pseudo-irrelevant update is a special case of an autonomously computable update.

\begin{note*}
By definition, if an update expression is irrelevant with respect to an extraction spanner, then it is also pseudo-irrelevant with respect to that spanner.
\end{note*}

\section{Categorizing Document Updates}\label{sec:Document Update categorization}
We wish to identify whether an update is irrelevant or pseudo-irrelevant with respect to a given extractor, independently of input documents. The essence of our approach is to inspect various kinds of \textit{overlap} between an update expression and an extractor. The proposed process verifies some \emph{sufficient} conditions for irrelevant, autonomously deletable, and pseudo-irrelevant updates.  

If an update changes the content length of an extracted span, then it will be relevant; the extractor should be re-executed.\footnote{There are some conditions under which the extracted relation after update might be autonomously computable. We leave the determination and detection of such conditions for future work.} 
However, even without changing an extracted value, an update could change the context for determining that a span should be extracted. 
First, updated spans, with the new value A, could form new matches for the extraction spanner, which would create new rows in the extracted view if we re-run the extractor. Second, some extracted spans might no longer match after the update, and therefore the associated rows would disappear when the extractor is re-run after the update.\footnote{These effects are not mutually exclusive.}

\begin{figure}[t]
 \centering
\includegraphics[width=0.5\columnwidth]{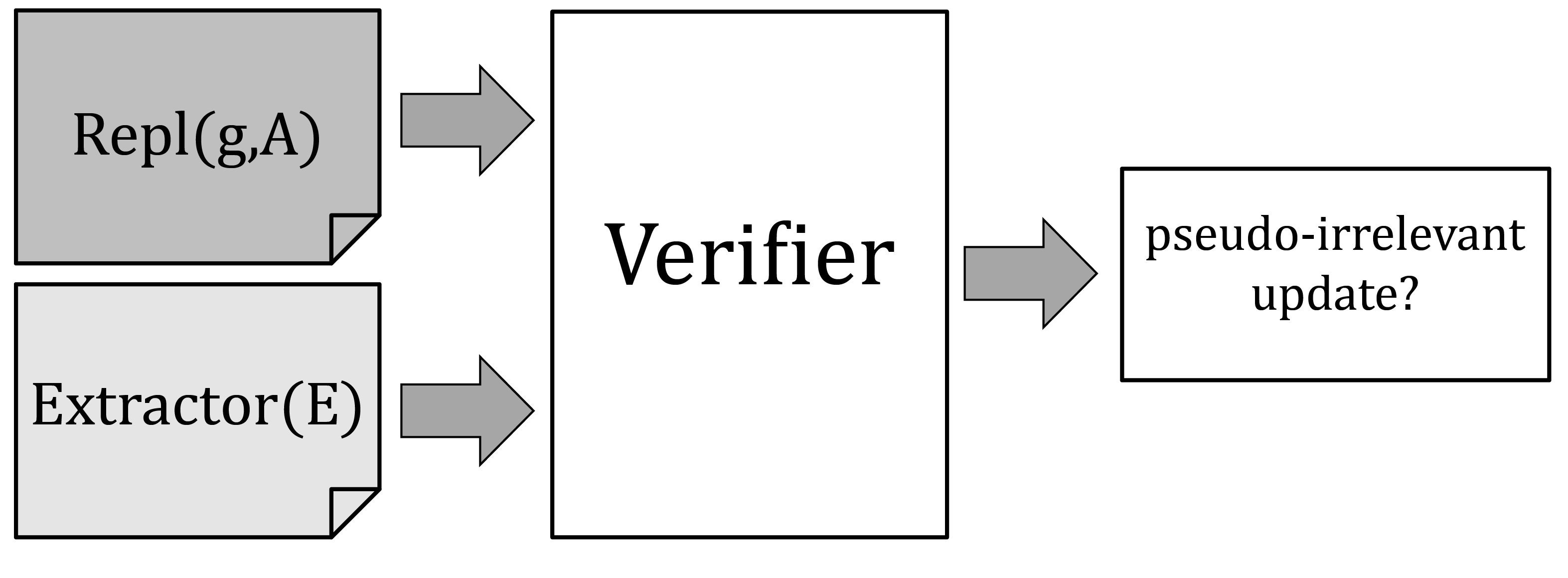}
\caption{The verifier statically analyzes an update expression and an extraction formula to test sufficient conditions for being a pseudo-irrelevant update.}  
\label{fig:verifier}
\end{figure}

After introducing a few simple constructs, we present a sound, but not necessarily complete, mechanism to determine whether an update expression, specified by the update formula $g$ and replacement string $A$, is pseudo-irrelevant with respect to a document spanner specified by an extraction formula $E$ (Figure~\ref{fig:verifier}).
\begin{definition}
Given $\mathit{Repl}(\gamma,A)$, the \emph{proxy language} $\nabla(g,A)$ is defined using the following disjunctive form: 
\[\nabla(g,A) = \bigvee_{i=1}^{k} V_i \]
where $V_i$ is derived from disjunct $U_i$ in $\Delta(g)$ by replacing the marked subexpression in that disjunct by $x\{A\}$, that is,
$
 V_{i}=\gamma^{\prime}_{L_{i}} \gdot x\{A\} \gdot \gamma^{\prime}_{R_{i}}
$
where $\gamma^{\prime}_{L_{i}}$ and $\gamma^{\prime}_{R_{i}}$ are the subexpressions preceding and following, respectively, the marked subexpression in $U_i$.
\end{definition}

We can now describe two simple special cases: 
\begin{enumerate} [(i)]
    \item  \label{itm:first} If $L(B(E)) \cap L(B(g))=\emptyset$, the update is irrelevant: there is no document on which both $E$ and $g$ match, and therefore any document that is updated cannot have extracted content.
    \item \label{itm:second} If  $L(B(E)) \cap L(B(g)) \neq \emptyset$ but  $L(B(E)) \cap L(B(\nabla(g,A))) =\emptyset$, there exist documents on which both $E$ and $g$ match, but if such a document is updated, the span relation produced by the extractor becomes empty. Although the update is relevant, it is autonomously computable: every extracted tuple from the updated relation is deleted.
\end{enumerate}

 We need to determine the relative positions of the capture variables in the extraction spanner and the unrestricted update spanner to determine whether an update is pseudo-irrelevant.

\subsection{Detecting Overlapping Spanners}
Clearly, if a document update changes some or all of the content of an extracted span, it will in general change the extracted span relation. 
Similarly, after an update, the replacement text might cause one or more additional spans to be extracted, so that the span relation includes tuples that did not meet the extraction condition before the update. 
We leave it to future work to determine under what conditions an update that overlaps extracted spans happens to be pseudo-irrelevant. Instead, we determine when there can be no overlap and then under which further conditions an update is pseudo-irrelevant.

\begin{definition}
Given two extraction formulas $E$ and $E'$, $\llbracket E \rrbracket$ and $\llbracket E' \rrbracket$ are \emph{disjoint} if for every document $D$, $\llbracket E \rrbracket(D)$ includes no span that overlaps with a span in $\llbracket E' \rrbracket(D)$.
Otherwise, we say that the spanners overlap.
\end{definition}

Given $\mathit{Repl}(g,A)$ and extractor $E$, we construct $\mathcal{M}_{\Bumpeq}$, to determine whether $\llbracket E \rrbracket$ and the unrestricted update spanner could produce at least one overlapping pair of spans, that is whether they could have at least one offset $o_2$ in common.
First we create a finite state machine $\mathcal{M}_{\vee }^{i}$ for each disjunct of $\Delta(E)$: 
$\mathcal{M}_{\vee }^{i}= \mathcal{M}_{R_0} \gdot \mathcal{M}{\gamma_{1}} \gdot \mathcal{M}_{R_1} \gdot \dots \gdot \mathcal{M}{\gamma_{n_i}} \gdot \mathcal{M}_{R_{n_i}}$
where $\mathcal{M}{\gamma_{m}}$ encodes the regular expression captured by the $m^{th}$ capture variable, and then we define $\mathcal{M}_{E}=<Q_{E},\Sigma_{E}, \delta _{E}, Q_{0_{E}}, F_{E}>$ by applying the standard union operator over $\mathcal{M}_{\vee }^{i}$ where $ 1 \leq i\leq k $. 
Next, we reuse  $\mathcal{M}_{g}=<Q_{g},\Sigma_{g}, \delta _{g}, Q_{0_{g}}, F_{g}>$ and the predicates $\mathcal{C}(Q,q,a)$ and $\mathcal{C}(Q_1,q,a,Q_2)$ that were introduced in Section~\ref{DisjointUpdateSpanner}.
Let $Q_{\gamma_m}$ denote the states in $\mathcal{M}{\gamma_m}$.
With these, we define $\mathcal{M}_{\Bumpeq}=<Q_{\Bumpeq},\Sigma_{\Bumpeq}, \delta _{\Bumpeq}, Q_{0_{\Bumpeq}},F_{\Bumpeq}>$ where
\begin{description}
\item[ ] $\Sigma_{\Bumpeq}=\Sigma_{\mathcal{M}_{g}} \cap \Sigma_{\mathcal{M}_{E}},$
\item[ ] $Q_{\Bumpeq}=Q_{g} \times Q_{E} \times \{T,F\},$
\item[ ] $Q_{0_{\Bumpeq}}=\{ (q_{i}, q_{j},F) \gbar q_{i} \in Q_{0_{g}} \wedge  q_{j} \in Q_{0_{E}}\},$
\item[ ] $F_{\Bumpeq}=\{ (q_{i}, q_{j},T) \gbar q_{i} \in F_{g} \wedge q_{j} \in F_{E} \},$ 
\item[ ] $\delta_{\Bumpeq}((q_{i} \times q_{j} \times v), a) =  
   \begin{dcases*} 
   (\delta_{g}({q_{i}},a),\delta_{E}({q_{j}},a),T) &if $(\mathcal{C}(Q_C,q_i,a) \vee \mathcal{C}(Q_L,q_i,a, Q_R)) \wedge$\\
   &$\exists k (\mathcal{C}(Q_{\gamma_{m}},q_j,a) \vee \mathcal{C}(Q_{R_{m-1}},q_j,a,Q_{R_{m}}))$ \\
   (\delta_{g}({q_{i}},a),\delta_{E}({q_{j}},a),v) & otherwise.
   \end{dcases*} $
\end{description}

\begin{proposition}\label{prop:disjointUpdate2}
  $L(\mathcal{M}_{\Bumpeq})=\{ D \gbar D $ is a witness for overlapping spans for update formula $g$ and extraction formula $E\}$.
\end{proposition}
\begin{proof}
The transition function identifies transitions that stay within a marked span or signal empty marked spans. The proof is then similar to that of Proposition~\ref{prop:disjointUpdate}.
\end{proof}
\begin{corollary} \label{cor:disjointupdate}
  Let $\llbracket g \rrbracket$ be an unrestricted update spanner specified  by update formula $g$ and $\llbracket E \rrbracket$ be a document spanner specified by extraction formula  $E$, and  construct automaton  $\mathcal{M}_{\Bumpeq}$ as above. If  $\mathit{min}(\mathcal{M}_{\Bumpeq}) = \emptyset$, $\llbracket g \rrbracket$ and $\llbracket E \rrbracket$  are disjoint. 
\end{corollary}	

Similarly, for a given extraction spanner and the proxy spanner for an update, we build a finite automaton, $\mathcal{M}^{p}_{\Bumpeq}$, to recognize the set of witnesses for overlapping spans. The  construction procedure is exactly the same as constructing $\mathcal{M}_{\Bumpeq}$, because a proxy spanner is isomorphic to a special case of an unrestricted update spanner with a constant string as the marked subexpression.    

\begin{proposition}\label{prop:disjointproxy}
  $L(\mathcal{M}^{p}_{\Bumpeq})=\{D | D $ is a witness for overlapping spans for the proxy formula $\nabla(g,A)$ and the extraction formula $E\}$.
\end{proposition}
\begin{proof}
Identical to  Proposition~\ref{prop:disjointUpdate2}.
\end{proof}
\begin{corollary} \label{cor:disjointproxy}
  Let $\llbracket \nabla(g,A) \rrbracket$ be a proxy spanner specified  by $\nabla(g,A)$ and $\llbracket E \rrbracket$ be a document spanner specified by extraction formula $E$,   construct automaton  $\mathcal{M}^{p}_{\Bumpeq}$ as above. If  $\mathit{min}(\mathcal{M}^{p}_{\Bumpeq}) = \emptyset$, $\llbracket \nabla(g,A) \rrbracket$ and $\llbracket E \rrbracket$  are disjoint. 
\end{corollary}	

\begin{theorem}\label{theorem:disjoint}
  For all documents, $\mathit{Repl}(g,A)$ is disjoint from $\llbracket E \rrbracket$ (i.e., $\llbracket g \rrbracket$ is disjoint from $\llbracket E \rrbracket$ and $\llbracket \nabla(g,A) \rrbracket$ is disjoint from $\llbracket E \rrbracket$) if  $\mathit{min}(\mathcal{M}_{\Bumpeq}) = \mathit{min}(\mathcal{M}^{p}_{\Bumpeq}) = \emptyset$ for automata  $\mathcal{M}_{\Bumpeq}$ and $\mathcal{M}^{p}_{\Bumpeq}$ as defined above. 
\end{theorem}
\begin{proof}
  This follows directly from Corollaries~\ref{cor:disjointupdate}~and~\ref{cor:disjointproxy}.
\end{proof}
\subsection{Detecting Pseudo-Irrelevance for Disjoint Spanners}

If an update is pseudo-irrelevant to an extractor, then all extracted spans must be shifted in a consistent manner. Therefore, the ordering within a document of the extracted spans forming each row in the extracted relation must remain unchanged after a pseudo-irrelevant update. Consider one disjunct from the normalized extraction formula for extraction (Lemma~\ref{lemma:disjFrom})
\[E_i = \theta_0 \gdot X_1\{\theta^\prime_1\} \gdot \theta_1 \gdot X_1\{\theta^\prime_2\} \gdot \theta_2 ... \gdot X_n\{\theta^\prime_n\} \gdot \theta_n\]
and a document $\alpha_0 \beta_1 \alpha_1 \beta_2 \alpha_2 ... \beta_n \alpha_n$ where $\alpha_j , \beta_j \in \Sigma^*$. If
$\alpha_j$ matches $\theta_j$ and $\beta_j$ matches $\theta^\prime_j$,
and if after substituting $A$ for strings identified by marked spans within the spans covering only the $\alpha_j$ the updated document still matches $E_i$, then the new locations of the spans matching $\beta_j$ will be simple shifts from their locations prior to the update. In fact, this will be true not only if the document matches the same $E_i$ after update, but also if it matches a disjunct that is similar to $E_i$ as defined here.

For each disjunct introduced in Lemma~\ref{lemma:disjFrom}, create a \emph{variable-profile} that expresses the relative position of each capture variable with respect to other variables. More specifically, given $E$ a formula matching the grammar for $\bar{\gamma}$, define $v(E)$ as the string produced from $E$ by eliminating all symbols except for capture variables and left and right braces. For example, if $E_1=s_{1} \gdot z\{s_2 \gdot y\{s_{6}\}\}\gdot x\{s_4\}$ where $s_i$ are (standard) regular expressions, then $v(E_1) = z\{y\{\}\}x\{\}$. Next, given an extraction formula $E$, let $\phi_E$ define a partitioning of the disjuncts in $\Delta(E)$ by their variable-profiles: \[\phi_E(E_i) = \{E_j  | E_j \mathit{~is~a~disjunct~in~}  \Delta(E) \mathit{~and~} v(E_i) = v(E_j)\}\]
\noindent where $E_i$ is a disjunct in $\Delta(E)$. Finally denote the union of all disjuncts in a partition as $\Phi_E(E_i) = \bigcup \phi_E(E_i)$ and let $\Phi(E) = \{\Phi_E(E_i)\;|\;E_i \in E\}$.

\begin{theorem}\label{theorem:resilience}
Given an update expression $\mathit{Repl}(g,A)$ defining an unrestricted update spanner and a disjoint extractor defined by $E$, let $E_i$ denote the disjuncts in $\Delta(E)$ and $L_i = L(B(\Phi_E(E_i)))$. 
The update is pseudo-irrelevant with respect to the extractor if and only if
\[
\forall D ([D = \mathit{Repl}(g,A)(D)] \gvee \forall E_i [D \in L_i \iff \mathit{Repl}(g,A)(D) \in L_i])
\]
\end{theorem}
\begin{proof}
(\textbf{only if:})  Assume that the update is pseudo-irrelevant with respect to the extractor. If $\forall D (D = \mathit{Repl}(g,A)(D))$ then the theorem holds. Otherwise, choose $D$ such that  $D' = \mathit{Repl}(g,A)(D) \neq D$. Let $D=\alpha_0 \beta_1 \alpha_1 \beta_2 \alpha_2 ... \beta_n \alpha_n$ where $\alpha_j , \beta_j \in \Sigma^*$ and each $\beta_i$ matches the capture variable in $\llbracket g \rrbracket$. (These must be non-overlapping.) Thus $D'=\alpha_0 A \alpha_1 A \alpha_2 ... A \alpha_n$. Because the extractor is disjoint from the update, all the extracted spans must appear within the $\alpha_j$ segments, and because the update is pseudo-irrelevant, the extractions from $D'$ must all be merely shifts from the extractions in $D$. But, in that case, the disjunct $E_{i_1}$ causing the extraction for $D$ must have the same variable profile as the disjunct $E_{i_2}$ causing the extraction for $D'$; that is, $\Phi_E(E_{i_1})= \Phi_E(E_{i_2})$ and the theorem holds.
 \\ (\textbf{if:}) Assume that the update is not pseudo-irrelevant with respect to the extractor. In that case, there exists a witness document $D$ such that applying $E$ to $D' = \mathit{Repl}(g,A)(D)$ produces a span relation where $\llbracket E \rrbracket (D') \neq \{S' \;|\; \exists\; S \in \llbracket E \rrbracket(D)$ such that $S'=\mathit{shift}(g,A)(S) \}$. That is, either (case 1) there is a span $s_1$ in $\llbracket E \rrbracket(D)$ that does not have a corresponding shifted span in $\llbracket E \rrbracket (D')$, or (case 2) there is a span $s_2$ in $\llbracket E \rrbracket(D')$ that is not simply a shift from some span in $\llbracket E \rrbracket (D)$.
Thus, $D \neq D'$.

\noindent \emph{Case if-1}: Let $E_i$ be a disjunct in $\Delta(E)$ that includes  $s_1$ in $\llbracket E_i \rrbracket (D)$\footnote{There must be such an $E_i$ because $s_1$ in $\llbracket E \rrbracket (D)$.} and thus $D \in L_i$. Let $E_i = \theta_0 \gdot X_1\{\theta^\prime_1\} \gdot \theta_1 \gdot X_1\{\theta^\prime_2\} \gdot \theta_2 ... \gdot X_n\{\theta^\prime_n\} \gdot \theta_n$
and $D = \alpha_0 \beta_1 \alpha_1 \beta_2 \alpha_2 ... \beta_n \alpha_n$ where $\alpha_j , \beta_j \in \Sigma^*$ and
$\alpha_j$ matches $\theta_j$ and $\beta_j$ matches $\theta^\prime_j$.

Because $D \neq D'$ and $\llbracket g \rrbracket$ does not overlap $\llbracket E\rrbracket$, there are some updates, all of which
must be replacements within $\alpha_0, \alpha_1, ..., \alpha_n$. Let $\alpha_i = \alpha_{i_0}s_{i_1}\alpha_{i_1}s_{i_2}\alpha_{i_2}...s_{i_{n_i}}\alpha_{i_{n_i}}$ where $s_{i_1}...s_{i_{n_i}}$ each match the capture variable in $\Delta(g)$. (Note that these matches must all be mutually disjoint because an unrestricted update spanner cannot produce overlapping spans.)
The update will replace $\alpha_i$ by $\alpha^\prime_i = \alpha_{i_0}A\alpha_{i_1}A\alpha_{i_2}...A\alpha_{i_{n_i}}$, producing the document $D' = \alpha^\prime_0 \beta_1 \alpha^\prime_1 \beta_2 \alpha^\prime_2 ... \beta_n \alpha^\prime_n$ and if each $\beta_j$ was within span $b_j$, then $b_j$ is shifted by $\mathit{shift}(g,A)(b_j)$.
Thus if the update is not pseudo-irrelevant, then clearly $D' \notin L_i$, because otherwise a disjunct with the same variable-profile as $E_i$ would match $D'$ and the shifted span would appear in the extracted relation for $D'$.

\noindent \emph{Case if-2}:
Let $E_i$ be a disjunct in $\Delta(E)$ that includes  $s_2$ in $\llbracket E_i \rrbracket (D')$ and thus $D' \in L_i$. Let 
$
E_i = \theta_0 \gdot X_1\{\theta^\prime_1\} \gdot \theta_1 \gdot X_1\{\theta^\prime_2\} \gdot \theta_2 ... \gdot X_n\{\theta^\prime_n\} \gdot \theta_n
$
and $D' = \alpha_0 \beta_1 \alpha_1 \beta_2 \alpha_2 ... \beta_n \alpha_n$ where $\alpha_j , \beta_j \in \Sigma^*$ and
$\alpha_j$ matches $\theta_j$ and $\beta_j$ matches $\theta^\prime_j$.

As before, if $D \neq D'$, some update occurred. Because $\llbracket \nabla(g,A) \rrbracket$ does not overlap $\llbracket E \rrbracket$,
all updates must have been replacements within $\alpha_0, \alpha_1, ..., \alpha_n$. Let $\alpha_i = \alpha_{i_0}A\alpha_{i_1}A\alpha_{i_2}...A\alpha_{i_{n_i}}$ where the indicated instances of the string $A$ are a result of the update (i.e., not already present in $D$). (Note that again these instances must all be mutually disjoint because an unrestricted update spanner cannot produce overlapping spans.)
The update will have created $\alpha_i$ from $\alpha^\prime_i = \alpha_{i_0}s_{i_1}\alpha_{i_1}s_{i_2}\alpha_{i_2}...s_{i_{n_i}}\alpha_{i_{n_i}}$ where $s_{i_1}...s_{i_{n_i}}$ match the capture variable in $\Delta(g)$. Thus $D = \alpha^\prime_0 \beta_1 \alpha^\prime_1 \beta_2 \alpha^\prime_2 ... \beta_n \alpha^\prime_n \in L(B(U_j))$ and if each $\beta_j$ was within span $b_j$ in $D$, then $b_j$ will have been shifted by $\mathit{shift}(g,A)(b_j)$.
Thus if the update is not pseudo-irrelevant, then clearly $D$ cannot be in $L_i$, because otherwise a disjunct with the same variable-profile as $E_i$ would match $D$ and the pre-shifted span would appear in the extracted relation for $D$.

Thus in both cases, 
$
\exists D \exists E_i \:(D \in L_i \iff 
\mathit{Repl}(g,A)(D) \notin L_i)\:)
$,
\noindent which completes the proof.
\end{proof}

Using this theorem, we construct a machine, i.e., $\mathcal{M}_R$, to recognize pseudo-irrelevant updates (Algorithm~\ref{alg:pseudo-detection}). Algorithm~\ref{alg:pseudo-detection} creates finite state machines using  standard operators including concatenation~($\gdot$),  union~($\cup$), intersection~($\cap$), and complement~($\overbar{M}$) \cite{DBLP:books/daglib/0016921}. (We assume that the built-in function $\mathit{fsm}()$ eliminates all capture variables from an input regular formula and converts the result to its equivalent finite state machine.)

\begin{figure}[ht]
\centering
\begin{minipage}{.92\textwidth}
\begin{algorithm}[H]
    \SetKwInput{Pre}{Precondition}
  	\KwIn{extraction formula $E$, \hspace{3in}update expression $\mathit{Repl}(g,A)$}
	\KwOut{automaton $\mathcal{M}_R$}
	\Pre{$\llbracket g \rrbracket$ unrestricted, \hspace{3in}$\mathit{Repl}(g,A)$ disjoint from $\llbracket E \rrbracket$}
    $\mathcal{M}_{\phi},\mathcal{M}_R  \gets \emptyset$\;
    \BlankLine
    \tcc{build extraction automata:}
    \For {$\Phi_i \in \Phi(E)$}
       {\tcc{$\Phi_i$ includes all disjuncts with the $i^{th}$ variable-profile} 
        $\mathcal{M}^{i}_{\phi} \gets \mathit{fsm}(\Phi_i)$\;
 	}
    \BlankLine
    \tcc{build document/update pairs:}
	\For {$ u_j \in \Delta(g)$ }
	{ $v_j \gets$ corresponding disjunct in $\nabla(g,A)$\; 
	  $\mathcal{M}^j \gets \mathit{fsm}(u_j) , \mathcal{M}^j_{\nabla} \gets \mathit{fsm}(v_j)$\;

    \ForAll{$\mathcal{M}^{i}_{\phi}$}
        { 	  $\mathcal{M}_R \gets \mathcal{M}_R \cup 
        ((\mathcal{M}^j \cap \mathcal{M}^{i}_{\phi}) \gdot (\mathcal{M}^j_{\nabla} \cap \overbar{\mathcal{M}}^{i}_{\phi}))$\;
         $\mathcal{M}_R \gets \mathcal{M}_R \cup 
        ((\mathcal{M}^j \cap \overbar{\mathcal{M}}^{i}_{\phi}) \gdot (\mathcal{M}^j_{\nabla} \cap \mathcal{M}^{i}_{\phi}))$\;
        }
    }    
    \BlankLine
	\Return{$\mathcal{M}_R$}
	\caption{Construction Algorithm for Recognizer of Pseudo-Irrelevant Updates.}
	\label{alg:pseudo-detection}
\end{algorithm} 
\end{minipage}
\end{figure}

\begin{proposition}
  $\mathit{min}(\mathcal{M}_R) \neq \emptyset$ iff $\exists$ a witness $D$ showing that the unrestricted spanner defined by $\mathit{Repl}(g,A)$ is not pseudo-irrelevant with respect to the disjoint extractor $\llbracket E \rrbracket$.
\end{proposition} 
\begin{proof}
First we prove that if there exists a witness document that shows  $\mathit{Repl}(g,A)$ is not  pseudo-irrelevant with respect to the  extractor $\llbracket E \rrbracket$ then $\mathit{min}(\mathcal{M}_R) \neq \emptyset$. Based on Theorem~\ref{theorem:resilience} if an update is not  pseudo-irrelevant  there exist  at least an input string $D$, an update disjunct $U_j$, and a partition $\Phi_E(E_i)$ such that\\ \emph{Case 1}: 
$D \in L_i \implies \mathit{Repl}(g,A)(D) \notin L_i \:$.
Based on the construction, the following holds:
\begin{description}
\item[ ] $D \in L(\mathcal{M}^j) \land \mathit{Repl}(g,A)(D) \in L(\mathcal{M}^j_\nabla)$
\item[ ] $D \in L(\mathcal{M}_{\phi}^i) \land \mathit{Repl}(g,A)(D) \in L(\overbar{\mathcal{M}}_{\phi}^i)$
\item[ ] $D \in L(\mathcal{M}^j \cap \mathcal{M}_{\phi}^i) \land \mathit{Repl}(g,A)(D) \in L(\mathcal{M}^j_\nabla \cap \overbar{\mathcal{M}}_{\phi}^i)$
\item[ ] $D \gdot \mathit{Repl}(g,A)(D) \in L((\mathcal{M}^j \cap \mathcal{M}_{\phi}^i) \gdot (\mathcal{M}^j_\nabla \cap \overbar{\mathcal{M}}_{\phi}^i))$
\item[ ] $D \gdot \mathit{Repl}(g,A)(D) \in L(\mathcal{M}_R)$
\item[ ] $\mathit{min}(\mathcal{M}_R) \neq \emptyset.$
\end{description}
\emph{Case 2}: 
$\mathit{Repl}(g,A)(D) \in L_i \implies D \notin L_i \:$.
Similarly to case 1:
\begin{description}
\item[ ] $D \gdot \mathit{Repl}(g,A)(D) \in L((\mathcal{M}^j \cap \overbar{\mathcal{M}}_{\phi}^i) \gdot (\mathcal{M}^j_\nabla \cap \mathcal{M}_{\phi}^i))$
\item[ ] $D \gdot \mathit{Repl}(g,A)(D) \in L(\mathcal{M}_R)$
\item[ ] $\mathit{min}(\mathcal{M}_R) \neq \emptyset.$
\end{description}
Next we show that if $\mathit{min}(\mathcal{M}_R) \neq \emptyset$ then there exists a witness $D$ that shows  $\mathit{Repl}(g,A)$ is not  pseudo-irrelevant with respect to the  extractor $\llbracket E \rrbracket$. Suppose $s \in  L(\mathcal{M}_R)$.

Then one of the iterations in the innermost loop of Algorithm~\ref{alg:pseudo-detection} must have inserted a term into $\mathcal{M}_R$. Thus, either $\exists i, j \mathrm{~s.t.~} s \in L((\mathcal{M}^j \cap \mathcal{M}_{\phi}^i) \gdot (\mathcal{M}^j_\nabla \cap \overbar{\mathcal{M}}_{\phi}^i))$ or $\exists i, j \mathrm{~s.t.~} s \in L((\mathcal{M}^j \cap \overbar{\mathcal{M}}_{\phi}^i) \gdot (\mathcal{M}^j_\nabla \cap \mathcal{M}_{\phi}^i))$. Thus $s$ is the concatenation of two strings $s_1$ and $s_2$ such that $s_1 \in L(\mathcal{M}^j)$ and $s_2 \in L(\mathcal{M}^j_\nabla)$. But then either $s_1 \in L(\mathcal{M}_{\phi}^i) \land s_2 \in L(\overbar{\mathcal{M}}_{\phi}^i)$ or $s_1 \in L(\overbar{\mathcal{M}}_{\phi}^i) \land s_2 \in L(\mathcal{M}_{\phi}^i)$. From this it follows that $D$ is a witness that $\mathit{Repl}(g,A)$ is not pseudo-irrelevant with respect to $\llbracket E \rrbracket$.
\end{proof}
\begin{corollary}\label{cor:pseudo}
If $\mathit{min}(\mathcal{M}_R)= \emptyset$ the update is pseudo-irrelevant.
\end{corollary}

From this, we arrive at a sufficient verification test for an update being pseudo-irrelevant with respect to an extractor as depicted earlier in Figure~\ref{fig:verifier}:

\begin{theorem}
Given an update expression $\mathit{Repl}(g,A)$ and a regular formula $E$, if $\mathit{min}(\mathcal{M}_{\Xi}) = \mathit{min}(\mathcal{M}_{\Bumpeq}) = \mathit{min}(\mathcal{M}^{p}_{\Bumpeq}) = \mathit{min}(\mathcal{M}_R) =  \emptyset$ for automata  $\mathcal{M}_{\Xi}, \mathcal{M}_{\Bumpeq}$, $\mathcal{M}^{p}_{\Bumpeq}$, and $\mathcal{M}_R$  as defined above, then the update expression is pseudo-irrelevant with respect to  $\llbracket E \rrbracket$.
\end{theorem}
\begin{proof}
Follows directly from Corollary~\ref{cor:unrestricted} ($\llbracket g \rrbracket$ is unrestricted), Theorem~\ref{theorem:disjoint} ($\mathit{Repl}(g,A)$ and $\llbracket E \rrbracket$ are disjoint), and Corollary~\ref{cor:pseudo} (updates must produce shifts).
\end{proof}

\section{Other Related Work}
\subsection{Information Extraction}
Expectations from extractors have risen as requirements have become more diversified, from the point that there were no  criteria to evaluate their performance~\cite{gaizauskas1998information} to the point that extraction algorithms need to work under various stresses such as noisy data, low response time, and diverse types of input and output~\cite{DBLP:journals/ftdb/Sarawagi08}.
The problems that deal with dynamic information sources are closest to our problem. These include continuous adaptation of extractors as their information sources changes and equipping  extractors with the ability to recycle previously obtained extraction results. For example, the approach by Lerman et al.~\cite{DBLP:journals/jair/LermanMK03} monitors updates on information sources for a specific class of extraction algorithms (wrappers) and rebuilds the extractor if the performance decreases due to the updates over their sources. In other work, Chen et al.~\cite{DBLP:conf/icde/ChenDYR08} efficiently update extractions when new documents are added to the source corpus: they identify  segments of new documents that have been seen previously by the extraction process and reuse their associated results.

\subsection{Document Spanners}
Researchers have addressed many problems using the document spanner model, including how to deal with documents with missing information~\cite{DBLP:conf/pods/MaturanaRV18} and  how to eliminate inconsistencies from extracted relations~\cite{DBLP:conf/pods/FaginKRV14}. Others have studied the complexity of evaluating spanners and computing the results of various algebraic operations over span relations~\cite{DBLP:conf/icdt/AmarilliBMN19, DBLP:journals/tods/FlorenzanoRUVV20, DBLP:conf/pods/PeterfreundFKK19, DBLP:conf/icdt/PeterfreundCFK19}.

In the presence of updates, the re-evaluation of an extractor might be sped up considerably if it is provably \emph{split-correct}, that is, if the extracted relation can be computed by combining the extractions from sub-documents~\cite{DBLP:conf/pods/DoleschalKMNN19}. Not only can extractions from various sub-documents then be run in parallel, but extractions can be completely avoided for those sub-documents that are not updated (i.e., those for which the update is irrelevant). 

Freydenberger and Thompson~\cite{DBLP:conf/icdt/FreydenbergerT20} have investigated the complexity of incrementally re-evaluating spanners in the presence of updates. However, their update model assumes that a document is encoded as a fixed-length \emph{word structure} in which (essentially) there is a special character that represents $\epsilon$ and the only operation is replacing one character from $\Sigma \cup \{\epsilon\}$ by another. 

In this work we have focused on a specific primitive representation for document spanners, i.e., so-called \textit{regex formulas}. It has been shown that the class of spanners defined by the more expressive \textit{variable-set automata} is closed under \textit{natural join} (as well as some other relational operators)~\cite{DBLP:journals/jacm/FaginKRV15, morciano2017engineering, DBLP:conf/pods/FreydenbergerKP18}, and this mechanism can be used to express various relationships between spans of a document~\cite{DBLP:journals/tods/FaginKRV16}. We plan to investigate whether variable-set automata can be adopted to simplify and extend our approach to determining pseudo-irrelevancy as well as other forms of autonomous updates.

\subsection{Static Analysis of Programs Using Regular Languages}
We use finite-state automata to determine whether an update expression is pseudo-irrelevant with respect to a document spanner. Similar static analyses of regular expressions have been used in diverse areas, including access control and feature interactions. For example, Murata et al.~\cite{DBLP:journals/tissec/MurataTKH06} propose an automaton-based, statically analyzed access control mechanism for XML database systems. 
In other work, an event-based framework is introduced for developing and maintaining new gestures that can be used in multi-touch environments~\cite{DBLP:conf/chi/KinHDA12}, 
and regular expressions associated with gestures are then statically analyzed to identify potential conflicts. 
Finally, we have also used finite automata to statically analyze extractors specified by JAPE~\cite{cunningham1999jape} in the context of updating extracted views~\cite{DBLP:conf/doceng/KassaieT19}.

\section{Conclusions}

\subsection{Summary of Main Results}
Perhaps our biggest contribution is the simple realization that information extraction can be considered as a view mechanism for document databases, subject to research similar to our community's vast experience with relational database views. The dual problems of efficiently maintaining materialized, possibly cascaded, views and of updating documents to reflect updates expressed against extracted views open up many opportunities for continued research that will ultimately lead to practical solutions.

This paper deals with the first of these problems only, and it provides a framework for exploring the basic ideas in extracted view maintenance. We have introduced a simple update model that can be applied to a document database and that is compatible with SystemT, a major extraction framework. We have begun to explore conditions for updates to be deemed irrelevant or autonomously computable with respect to extractors defined using that framework. 
Finally, we have described a particular form of autonomously computable update, namely pseudo-irrelevance, we have determined sufficient conditions for an update to be pseudo-irrelevant, and we have designed automata to test those conditions for given update expressions and extractors.

\subsection{Future Work}
We have established some sufficient conditions for updates to be pseudo-irrelevant, but we have not yet investigated whether there are necessary conditions as well. Furthermore, we have not yet investigated other autonomously computable conditions, such as those that might result in span modifications or insertions of extracted tuples. We have also not yet explored the practicality of constructing our verification automata nor investigated update properties of extractors that are defined by mechanisms more expressive than spanners.

Our model for document updates is also quite limited.  First of all, only one variable is used to identify spans that can be updated, even though correlated updates might require multiple related variables to update. Secondly, the substitute value is limited to being a constant, whereas real world applications might need to use various values based on some factors, such as the relative position of the update, some associated string values, or the contexts of matched spans.  Thirdly, for each document, all intended spans are updated once and simultaneously, a fundamental assumption that can be violated in practical situations. 
Loosening any of these restrictions creates new research challenges for verifying pseudo-irrelevance or other update properties.

\section*{Acknowledgements}

We gratefully acknowledge financial assistance received from the University of Waterloo and NSERC, the Natural Sciences and Engineering Research Council of Canada.

\bibliography{ExtractedView}

\end{document}